\documentclass[%
superscriptaddress,
 amsmath,amssymb,
pra,
twocolumn
]{revtex4-1}

\usepackage{graphicx}
\usepackage{dcolumn}
\usepackage{bm}

\usepackage {graphicx}

\usepackage{float}

\usepackage{pgfplots}
\pgfplotsset{width=7cm,compat=newest}

\usepackage[utf8]{inputenc}

\usepackage []{csquotes}

\usepackage{tikz}

\usepackage{amsthm}

\usepackage{amstext}

\usepackage {amsmath}

\usepackage{bbm}

\usepackage{bbold}

\usepackage{accents}

\usepackage{lettrine}

\usepackage[]{hyperref}

\definecolor{rosso}{RGB}{200,0,0}

\renewcommand{\vr}{\mathbf{r}}

\newcommand{\vk}{\mathbf{k}}
\newcommand{\vq}{\mathbf{q}}

\newcommand{\be}{\begin{equation}}
\newcommand{\ee}{\end{equation}}

{\left\lbrace\begin{array}{@{}l@{}}}%
{\end{array}\right.}

\begin{document}


\title{Breaking of Goldstone modes in two component Bose-Einstein condensate}

\author{Alessio Recati}
\affiliation{Arnold Sommerfeld Center for Theoretical Physics,
Ludwig-Maximilians-Universit\"at M\"unchen, 80333 M\"unchen, Germany}
\affiliation{INO-CNR BEC Center and Dipartimento di Fisica, Universit\`a di Trento, 38123 Povo, Italy}
\author{Francesco Piazza}
\affiliation{Institut f\"ur Theoretische Physik, Universit{\"a}t Innsbruck, A-6020~Innsbruck, Austria}

\date{\today}

\begin{abstract}
We study the decay rate  $\Gamma(k)$ of density excitations of
two-component Bose-Einstein condensates at zero temperature. Those
excitations, where the two components oscillate in phase, include 
the Goldstone mode resulting from condensation.
While within Bogoliubov approximation the density sector and the spin
(out-of-phase) sector are independent, they couple at the three-phonon
level. For a Bose-Bose mixture we find that the Belyaev decay is slightly modified due to the coupling with the 
gapless spin mode. At the phase separation point the decay rate changes instead from the standard $k^5$ to a $k^{5/2}$ 
behaviour due to the parabolic nature of the spin mode. 
In presence of coherent coupling between the two components the spin
sector is gapped and, away from the ferromagnetic-like phase
transition point, the decay of density mode is not affected. 
On the other hand at the transition point, when the spin fluctuations become critical, 
the Goldstone mode is not well defined anymore since $\Gamma(k)\propto k$. 
As a consequence, we show that the
friction induced by a moving impurity is enahnced -- a feature which
could be experimentally tested.
Our results apply to every non-linear 2-component quantum hydrodynamic Hamiltonian
which is time-reversal invariant, and possesses an $U(1)\times {\mathbf Z}_2$ symmetry.

\end{abstract}

\maketitle


\section{Introduction}

The existence of Goldstone modes \cite{Goldstone1961}, i.e. gapless
collective excitations, has crucial consequences on the thermodynamics
and dynamics of systems with spontaneously broken continuous
symmetries. While expected to be generically present in such systems, they can actually disappear in some specific 
situations. The most famous is the Anderson-Higgs mechanism \cite{Englert1964,Higgs1964}, 
known in the relativistic context where  for instance a scalar Higgs field gives a finite mass to the W- and Z-Bosons 
in electroweak theory, i.e. three out of the four Goldstone modes associated with the four generators of $U(1) \times SU(2)$ 
become massive. This effect can be understood as due to the long-range interactions and is present also in non-relativistic systems like superconductors  \cite{Anderson1958} - where the phase mode characterising cooper-pair condensation disappears and the photons become massive - or jellium \cite{Wigner1938} - where the Wigner crystal loses one of the three goldstone modes corresponding to translational symmetry breaking.

Here we introduce a new scenario for the breaking of the Goldstone modes, where the latter do not become massive but rather acquire a fast decay channel making them not well defined excitations. This happens due to the coupling of the Goldstone modes with further gapless collective modes into which they can decay, the latter appearing due to the spontaneous breaking of a further discrete symmetry. 
This mechanism carries analogies with the one predicted for systems possessing a Fermi surface \cite{DasSarma2015}, the latter indeed showing gapless single-particle excitations into which the Goldstone modes can decay.

Our system consists of a two-component weakly-interacting
Bose-Einstein condensate (BEC) whose internal levels are coherently
driven by an external electromagnetic field. The system shows both
density (in-phase) and spin (out-of-phase) collective excitations
\cite{Goldstein1997,Blakie1999,Search2001,Tommasini2003}. The former
are the $U(1)$ gapless phonons characterising the condensation, while
the latter are gapped and they become gapless at a ferromagnetic critical point for the
spontaneous breaking of the $\mathbf{Z}_2$ symmetry corresponding to
the exchange of the two components. The vanishing of the gap makes the
density modes decay into two spin modes with a rate of the same order
of their energy, i.e. the density modes become not well defined
excitations. This implies for instance that a moving impurity would
generate an enhanced friction, which we compute analytically.

Our results are more general than the two-component BEC studied
here. They would namely apply to any non-linear quantum hydrodynamic time-reversal-invariant
Hamiltonian which couples density and spin, possessing an $U(1)\times {\mathbf Z}_2$ symmetry.

We also consider the case without the interconversion term, also known as a Bose-Bose mixture, 
which posses a $U(1)\times U(1)\times{\mathbf Z}_2$ symmetry. Both the density and the spin excitations 
are gapless and linear.
The system phase separates when the spin compressibility (susceptibility) diverges. Although enahnanced the 
decay rate of density modes scales, in this case, at a slower rate than thier energy.

\section{Model}

We consider an atomic Bose gas at zero temperature, whose atoms of mass $m$ have two internal levels $|a\rangle$ and $|b\rangle$. The latter are typically magnetically trappable hyperfine levels. An external field is applied that couples the $|a\rangle$ to the $|b\rangle$ state via usually a two-photon transition, characterised by a Rabi splitting $\Omega$ that we take real and positive.  
The atoms interact via short range interactions described by the
strengths, $g_{aa}$, $g_{bb}$ and $g_{ab}$ corresponding to the intra-
and the inter-species collisions, respectively. Introducing the fields
$\hat\psi_j$, with $j=a$, $b$ the microscopic Hamiltonian can be
written as
\begin{eqnarray}
\label{Hmicro}
H&=&\int d\vr\left[\sum_{j=a,b}{\hbar^2\over 2m}|\nabla\hat\psi_{j}|^2+\sum_{i,j}\frac{g_{ij}}{2}\hat\psi^\dagger_i\hat\psi^\dagger_j\hat\psi_j\hat\psi_i\right]\nonumber\\
&&+\int d\vr\;{\hbar\Omega\over 2}(\hat\psi^\dagger_a\hat\psi_b+\hat\psi^\dagger_b\hat\psi_a).
\end{eqnarray}

The system has an $U(1)$ symmetry for $\Omega\neq 0$, corresponding to the total atom number being conserved, and an $U(1) \times U(1)$ symmetry for $\Omega=0$, corresponding to both total and relative particle numbers being conserved. 
At  $T=0$ the system is a Bose-Einstein condensate (BEC) described by the complex spinor order parameter 
$(\Psi_a(\vr,t),\Psi_b(\vr,t))$, where $\Psi_j$, $j\in\{a,b\}$ is the wave function macroscopically occupied by atoms in the internal state $|j\rangle$. 
For the sake of clarity we consider $g_{aa}=g_{bb}\equiv g$ in which case the system posses a further ${\mathbf Z}_2$ symmetry, corresponding to the exchange of the two components.

Introducing the amplitude and phase representation $\Psi_j=\sqrt{n_i}\exp(i\phi_j)$ the mean-field energy functional reads
\begin{eqnarray}
\label{MFH}
E_{MF}&=&\sum_{j=a,b}\int d\vr\left({\hbar^2\over 2m}|\nabla\sqrt{n_j}|^2+{\hbar^2 n_j\over 2m}|\nabla\phi_j|^2 +{1\over 2}gn_j^2 \right)\nonumber\\
&&+\int d\vr \left(g_{ab}n_a n_b+\hbar\Omega\sqrt{n_an_b}\cos(\phi_a-\phi_b)\right).
\end{eqnarray}

\begin{figure}
\includegraphics[width=0.5 \textwidth]{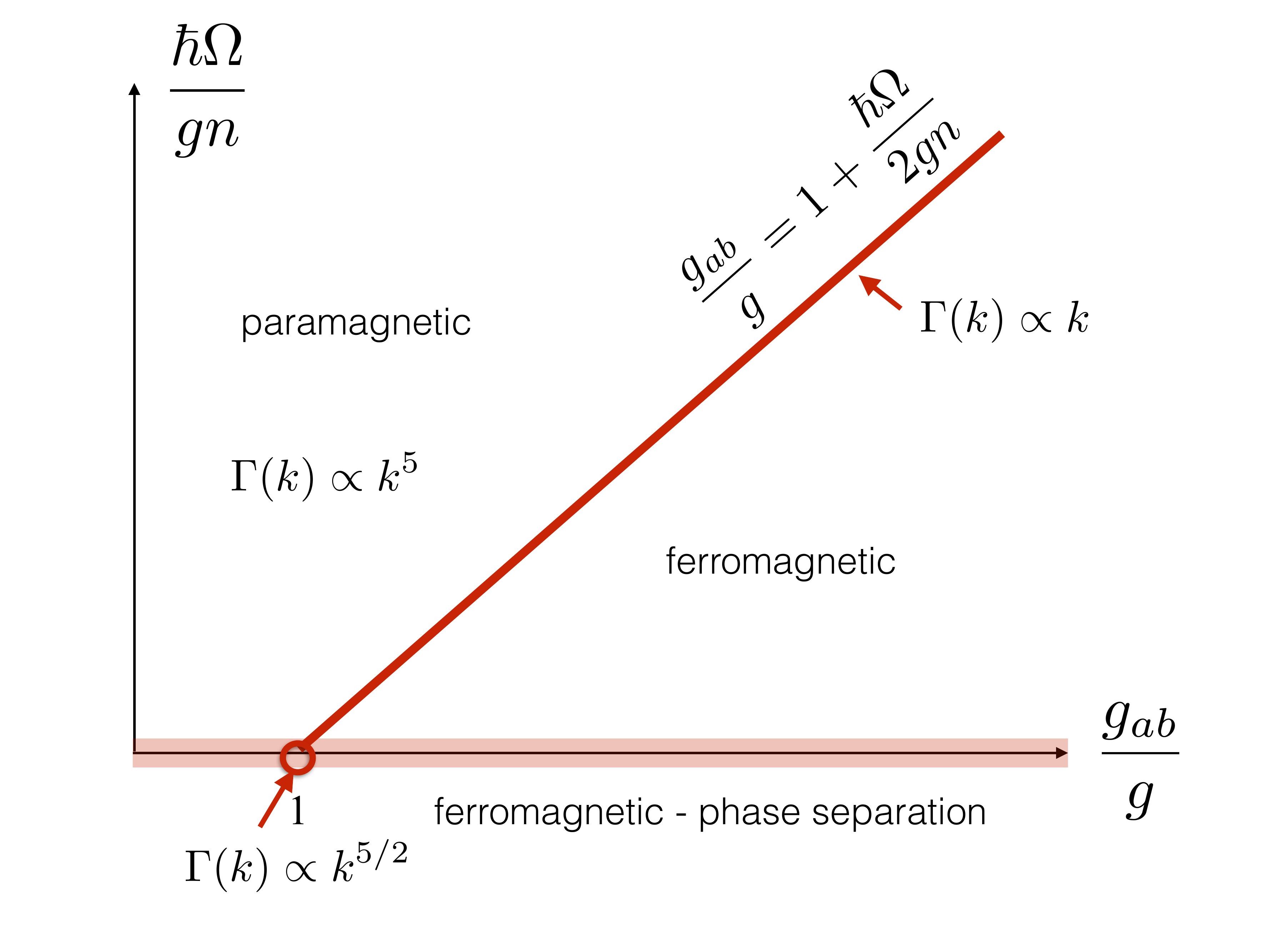}
\caption{Sketch of the phase diagram of two component Bose-Einstein condensates with density $n$ in presence of both 
intra- and inter-species interactions, $g$ and $g_{ab}$, respectively, as well as a coherent interconversion term $\Omega$ 
between the two species. The system exhibits a ferromagnetic-like phase transtition for strong enough interspecies interaction.
For $\Omega=0$ since the total magnetization is preserved the transition has a different character with respect to the 
$\Omega\neq 0$ case. In particular, Belyaev decay $\Gamma(k)$ strongly differs in the two cases (see text).}
\label{fig:phasediag}
\end{figure}

The ground state of the system is homogeneous with a fixed relative
phase $\phi^0_a-\phi^0_b=\pi$  -- due to the last term in
Eq. (\ref{MFH}) -- and, as already mentioned, can be either an
unpolarised paramagnetic phase with $n^0_a=n^0_b=n$ or a partially
polarised ferromagnetic phase $n^0_a\neq n^0_b$, which breaks the
$\mathbf{Z}_2$ symmetry. The transition between the two phases is
second order and occurs for $\hbar\Omega=\hbar\Omega_c=(g_{ab}-g)n$
(see, e.g., Ref.~\onlinecite{marta2013} and reference therein). The phase
transition between the unpolarised and polarised phase has been experimentally observed in Ref.~\onlinecite{Zibold2010}.   
A sketch of the phase diagram is reported in Fig. \ref{fig:phasediag}, where the singular nature of the $\Omega=0$ ferromagnetic transition is also put in evidence.

Above the ground state coherently coupled two-component Bose gases have two excitations branches: a gapless density or in-phase mode, which is the Goldstone mode related to the symmetry $U(1)$, and a gapped spin or out-of-phase mode, which becomes gapless at zero momentum at the ferromagnetic transition point.  

We derive the known results within a quantum hydrodynamic formalism for the paramagnetic phase in order to fix the notation
we need in the rest of the paper. 
We introduce the fluctuation fields $\Pi_j$ and $\phi_j$, $j=a,\;b$ for the amplitude and phase, respectively, 
and their in-phase (density)  $\Pi_d=(\Pi_a+\Pi_b)/2$, 
$\phi_d=\phi_a+\phi_b$ and out-of-phase (spin) $\Pi_s=(\Pi_a-\Pi_b)/2$, $\phi_s=\phi_a-\phi_b$ linear combinations.

In this way, the non-linear quantum hydrodynamic Hamiltonian obtained by expanding the Hamiltonian Eq. (\ref{MFH}) 
to quadratic order in the fluctuation fields decomposes in two sectors $H^{(2)}=H^{(2)}_d+H^{(2)}_s$, where 
\begin{eqnarray}
 \label{2order}
H^{(2)}_d\!\!&=&\!\!\int\!\! d\vr \!\!\left[{\hbar^2|\nabla\Pi_d|^2\over 4mn}+g_d\Pi_d^2+{\hbar^2n|\nabla\phi_d|^2\over 4m}\right],\\
H^{(2)}_s\!\!&=&\!\!\int\!\! d\vr \!\!\left[{\hbar^2|\nabla\Pi_s|^2 \over 4mn}+g_s\Pi_s^2+{\hbar^2n|\nabla\phi_s|^2\over 4m}+{\hbar\Omega n\over 2}\phi_s^2\right].
\label{H2spin}\end{eqnarray}
In the above equations we introduced the coupling constants $g_d=g+g_{ab}$ and $g_s(\Omega)=g-g_{ab}+\hbar\Omega/2n$. The quadratic Hamiltonian Eq. (\ref{2order}) can be easily diagonalised by introducing  the annihilation (creation) operators for the density $d_\vk$ ($d^\dagger_\vk$) and spin mode $s_\vk$ ($s^\dagger_\vk$) at momentum $\vk$ as
\begin{eqnarray}
\Pi_\alpha(\vr)&=&\sqrt{n\over 2}\sum_\vk { U}_{\alpha,k}(\alpha_\vk e^{i\vk\cdot\vr}+\alpha^\dagger_\vk e^{-i\vk\cdot\vr}), 
\label{Pi}
\\
\phi_\alpha(\vr)&=&i\sqrt{1\over 2n}\sum_\vk{U}_{\alpha,k}^{-1}(\alpha_\vk e^{i\vk\cdot\vr}-\alpha^\dagger_\vk e^{-i\vk\cdot \vr}),
\label{fi}
\end{eqnarray}
with $\alpha=d,\,s$ and where we defined (see also Ref.~\onlinecite{Tommasini2003} for the most general case $g_a\neq g_b$)
\begin{equation}
U_{d,k}=\left({k^2\over k^2+4mg_dn}\right)^{{1\over 4}},
U_{s,k}=\left({k^2+2m\hbar\Omega \over k^2+4mg_sn}\right)^{{1\over 4}}.
\label{Us}
\end{equation}
The density and spin Hamiltonians now simply read 
\begin{eqnarray}
 \label{2orderDiag}
H^{(2)}_d\!\!&=&\sum_\vk \omega^d_\vk d^\dagger_\vk d_\vk ,\;\omega^d_\vk=\!\!\sqrt{{\hbar^2k^2\over 2m}\left({\hbar^2k^2\over 2m}+2g_d n\right)}\\
H^{(2)}_s\!\!&=&\sum_\vk \omega^s_\vk s^\dagger_\vk s_\vk ,\; \omega^s_\vk=\!\!\sqrt{\left({\hbar^2k^2\over 2m}+2\hbar\Omega\right)\!\!\left({\hbar^2k^2\over 2m}+2g_s n\right)}\nonumber
\end{eqnarray}
Therefore, while the density mode is gapless and linear at small momenta, the spin mode has a gap 
$\Delta_s=2\sqrt{\hbar\Omega g_s n}$. 

From the previous analysis the difference between a mixture, $\Omega=0$, 
and the case $\Omega\neq 0$ is very clear.
For $\Omega=0$ the density and the spin sector behave in the same
way. The spectra are both gapless and the low momentum excitations are
phase-like, as it has to be for Goldstone modes of the $U(1)\times U(1)$ broken symmetries. 
The stiffnesses of the density and the spin modes are related to $g_d=g+g_{ab}$, $g_s(0)=g-g_{ab}$. 
On the verge of phase separation, i.e., $g_s(0)=0$, the spin mode becomes quadratic at low momenta and it acquires 
an amplitude contribution, being now both the relative phase and the relative amplitude fluctuations finite at low momenta.

On the other hand for $\Omega\neq 0$  at the transition point,
$g_s(\Omega_c)=0$, the gap closes, the low energy spin-mode is linear
and dominated by relative amplitude fluctuations $\Pi_s$ as it is clear already from Eq. (\ref{H2spin}). The latter become critical since the instability is due to the system breaking ${\mathbf Z}_2$ and building a finite polarisation.

\section{Belyaev decay for two-component Bose gas}

At the Bogoliubov level the modes are well defined. Finite lifetime comes by including higher order terms which represent interaction among various modes.
In particular, the third order term represents the so-called Belyaev
decay of one excitation into two new excitations and is
the dominant process at low temperatures\cite{LL9} . 
In a single component weakly interacting Bose gas the decay rate $\Gamma$ of phonons at low momentum $\vk$ is very small 
$\Gamma(\vk)\propto k^5$ (see also Table \ref{tabsum}).

In the case of a 2-component Bose gas further decay processes are in principle possible since, e.g., 
a density mode can decay into two spin modes. 
At the phase transition point the spin modes change their
character. We show in the following that this leads to a strong
enhancement of the Belyaev decay rate. In particular, we anticipate
here (see also Table \ref{tabsum}) that the Goldstone mode is still
well defined for a mixture $\Omega=0$ with a decay rate which scales
like $k^{5/2}$, while for $\Omega\neq 0$ the Goldstone mode is not properly defined, 
since the decay rate scale like its energy, i.e., $\Gamma(\vk)\propto k$.

\subsection{Symmetries and the general structure of the three-mode vertices}

\begin{figure}
\includegraphics[width=0.45 \textwidth]{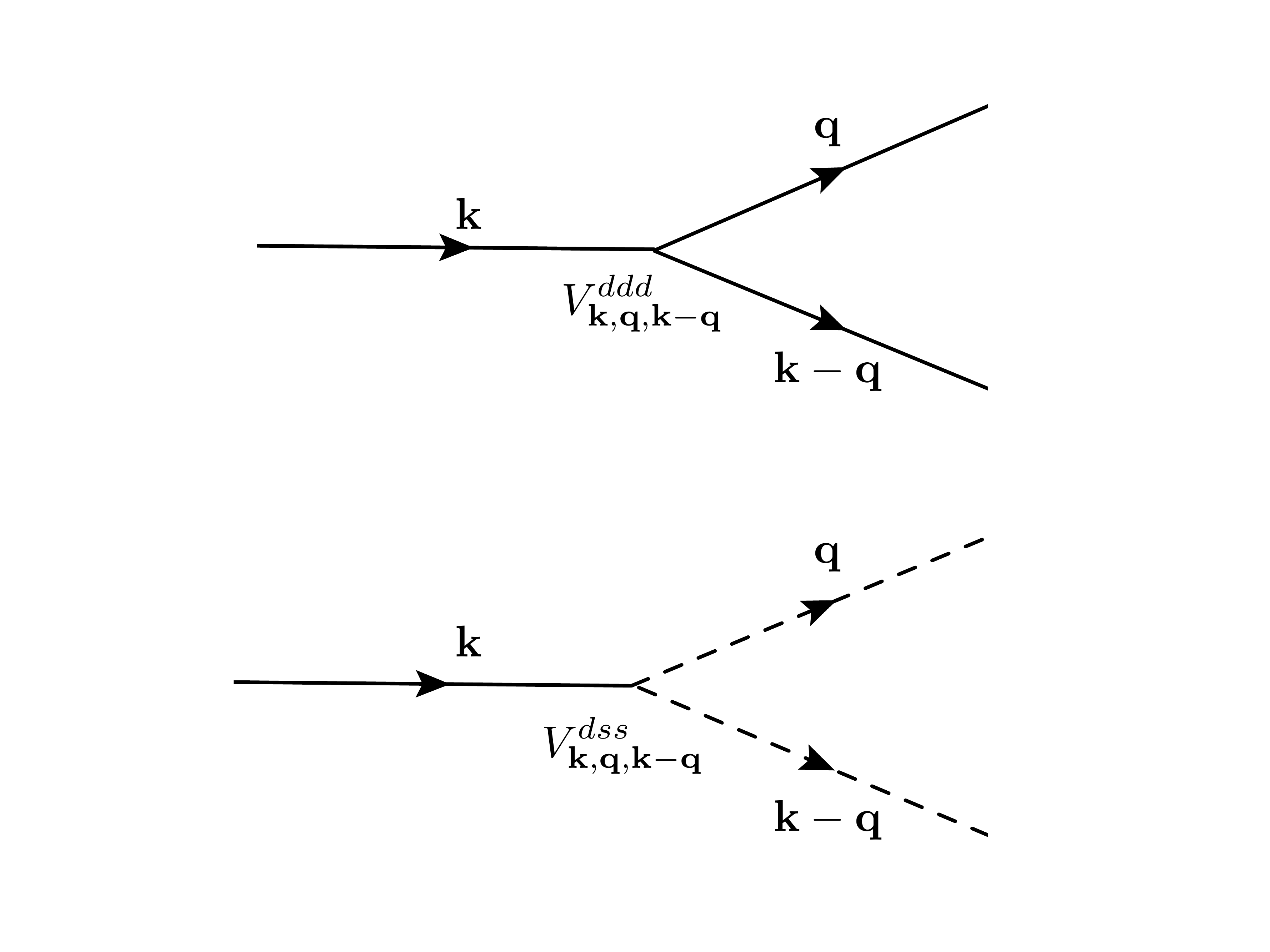}
\caption{Possible three mode vertices}
\label{fig:loops}
\end{figure}

To obtain the vertices of the possible decay processes we have to expand Eq. (\ref{MFH}) to third order.
The number of non-zero terms is pretty small due to the symmetries of the system.
In the paramagnetic phase due to the ${\mathbf Z}_2$ symmetry all the terms with an odd number of 
spin fields have to be zero. Therefore, the density mode can decay either in (i) two density modes or in 
(ii) two spin modes, as schematically represented in Fig. (\ref{fig:loops}). 
Moreover, due to the total density $U(1)$ symmetry the process (i) can occur only via 
$\Pi_d|\nabla\Pi_d|^2$, $\Pi_d^3$ and $\Pi_d|\nabla\phi_d|^2$, which lead to the standard Belyaev decay. 
The possible terms related to process (ii) are $\Pi_d|\nabla\phi_s|^2$ and  $\Pi_s\nabla\phi_d\nabla\phi_s$ for $\Omega=0$, 
while also the terms $\Pi_d\Pi_s^2$ and  $\Pi_d\phi_s^2$ are present for $\Omega\neq 0$. 
For instance, the term $\Pi_d\Pi_s^2$ gives rise to the following vertex:
\begin{equation}
-{\Omega\over 2n^2 }U_{d,\vq_1}U_{s,\vq_2}U_{s,\vq_3}
\end{equation} 
As we show below, such a vertex is responsible for the breaking of the Goldstone mode at 
the critical point for the ferromagnetic-like transition. 

\subsection{Results}

The decay rate is given by the imaginary part of the self-energy for the density mode. We calculate the self-energy  
at the one-loop level, which coincides with a Fermi's golden rule calculation. 
The general expression for one of the above mentioned process reads
\be
\Gamma(\vk)={\mathtt{Pm}_V\over (2\pi)^2}\int d^3{\vq} |V_{\vk,\vq,\vk-\vq}|^2
\delta(\omega^d_k-\omega^s_q-\omega^s_{|\vk-\vq|})
\ee
where $V$ is the vertex of the process and $\mathtt{Pm}_V$ the number of possible equivalent diagrams.

Since we are interested in the decay rate at low momentum we can consider only the most relevant terms in the different 
regimes as reported in Table \ref{tabsum}. For completeness we also report the result for a single-component Bose gas and 
for different dimensionality.
\begin{table}[htdp]
\caption{Belyaev decay of the density Bogoliubov mode}
\begin{center}
\begin{tabular}{|c|c|c|}
system & $\Gamma(\vk)$ (D=dimension) & dominant term\\
\hline
1-comp. Bose gas & 0 (D=1), $k^{2D-1}$ (D$>1$) & \\
2-comp. Bose gas & $k^2$ (D=1), $k^{2D-1}$ (D$>1$) & \\
$\Omega=0$ PS point & $k^{D/2+1}$ & $\Pi_s\nabla\phi_s\nabla\phi_d$\\
$\Omega\neq 0$ FM transition & $k^{D-2}$ & $\Pi_d\Pi_s^2$\\
\end{tabular}
\end{center}
\label{tabsum}
\end{table}%
\paragraph{Mixture $\Omega=0$.}

In the case of a mixture $\Omega= 0$ and away from the phase separation $g\neq g_{12}$ one has the ordinary Belyaev decay, where the pre-factor is renormalised due to the decay of density in two spin phonons. 
The most relevant terms at low momentum are the $\Pi_d(\nabla\phi_d)^2$ for the three density phonon vertex and $\Pi_d(\nabla\phi_s)^2$ and $\Pi_s\nabla\phi_s\nabla\phi_d$ for the density into two spin phonon vertex.
The decay rate reads 
\be
\Gamma(\vk)\simeq {3 k^5 (1+h(c_d/c_s))\over 640 nm\pi},
\ee
where $h(r)=7r/12+43/72 r-11r^3/24+5r^5/18$, which for two non-interacting species reduces to $h(1)=1$.
At the phase separation point  the most relevant term is only  $\Pi_s\nabla\phi_s\nabla\phi_d$ as can be seen by putting $\Omega$ and $g_s$ to zero in Eq. (\ref{Us}) and one gets a strong enhancement of the phonon decay which now reads
\be
\Gamma(\vk)={(mc_d k)^{5/2}\over 48 nm\pi}.
\ee
Still the density phonon mode is well defined at low momenta since $\Gamma_k/\omega_k\simeq k^{3/2}\rightarrow 0$.

\paragraph{Coherent coupling $\Omega\neq 0$.}
 
When the coherent coupling is on, the spin sector is gapped, therefore away from the transition point and at zero temperature it does not contribute to the phonon decay which is simple due the standard 
Belyaev process $\Pi_d(\nabla\phi_d)^2$, leading to $\Gamma(\vk)={3 k^5/(640 nm\pi)}$.

At the ferromagnetic transition the situation is very different. The gap in the spin channel closes and the spectrum  becomes linear at small momentum, i.e., $\omega_s(k)=c_s|k|$ with $mc_s^2=(g_{12}-g)n=\Omega_c$ where $\Omega_c $ is the value of the coherent coupling at the transition point. A density phonon can now decay into two spin ones. The latter are critical at the transition and, as already mentioned, dominated by the relative amplitude fluctuations, since the system is on the verge of polarization. The most relevant term becomes $\Pi_d\Pi_s^2$, whose contribution leads to a critical decay rate
\be
\Gamma(\vk)={(mc_s)^4 k\over 4 nm\pi},
\ee
making the Goldstone mode a not well defined excitation.

The decay rate of the density excitations can be measured having access to the dynamic structure factor 
$S(\vk,\omega)$. In the field of cold gases an accurate measurement of $S(\vk,\omega)$ is difficult. The measurement is 
based on Bragg spectroscopy and it has been used mainly to extract the resonance energies \cite{KetterleBragg,SteinhauerSpec}. 
However recently a new promising method has been demonstrated by coupling the gas with the mode of an high-finesse cavity 
\cite{EsslingerSkw}. 

An indirect effect of the short lifetime of the phonons is instead the response of the system to a local density perturbation 
as we describe in the following section.
 
\section{Force on an impurity - Friction}

Landau theory of superfluidity leads to the existence of a finite
critical velocity below which the flow is dissipationless. A moving
object weakly interacting with the fluid feels a friction force only
if its speed is larger than the Landau critical velocity. For
homogeneous ultra-cold gases the situation is quite clear and the
critical velocity is due to Cherenkov phonon emission \cite{PhysRevLett.97.260403}. If phonons have a finite life-time a friction force is present for any speed of the moving impurity.

The dissipation of energy due to a time-dependent potential can be
generally written in terms of the dynamic structure factor $S(\vk,\omega)$ as 
\be
\dot E=-\int \frac{d \vk}{(2\pi)^3}\int_0^\infty \frac{d\omega}{2\pi}\omega S(\vk,\omega)|W(\vk,\omega)|^2,
\ee
where $W(\vk,\omega)$ is the Fourier transform of the external perturbation.
Considering a delta-like  infinite mass impurity moving at a constant speed $\bf V$, we can write $W(\vr,t)=\lambda\delta({\vr}-{\mathbf V}t)$ where $\lambda$ is the coupling between the impurity and the gas, which leads to $W(q,\omega)=2\pi\lambda\delta(\omega -{\mathbf q}\cdot{\mathbf V})$. 

Accounting for the finite phonon lifetime $\Gamma(\vk)$ at the on-shell level corresponds in writing the dynamic structure factor as
\be
S(\vk,\omega)=n|U_d(\vk)|^2{\Gamma(\vk)\over (\omega-\omega^d_\vk)^2+\Gamma(\vk)^2 }.
\ee
Therefore, the expression for dissipated energy per unit time reads
\be
P=\frac{2\pi}{\hbar}\lambda^2\int \frac{d\vk}{(2\pi)^3}n|U_d(\vk)|^2{\Gamma(\vk)\over (\vk\cdot {\bf V}-\omega^d_\vk)^2+\Gamma(\vk)^2 }\vk \cdot{\bf V}.
\ee
Considering that at low speed $|\bf V|$ the most relevant contribution comes from momenta $k<{\bar k}\ll 1/\xi_d$ with  $\xi_d=\hbar/mc_d$ the density healing length, we find that the dissipated energy depends quadratically on the speed of the impurity and scale very differently far from the transition and at the transition point, namely
\begin{equation}
P=-\frac{\lambda^2 }{12\pi^2\xi_d^6}\left(\frac{V}{c_d}\right)^2\left \{
  \begin{tabular}{cc}
   $\frac{3}{160 }({\bar k}\xi_d)^8$, & $\Omega>\Omega_c$ \vspace{0.2cm}\\
   $\frac{(c_s c_d)^4}{(c_d^4+c_s^4)^2}({\bar k}\xi_d)^4$, & $ \Omega=\Omega_c$
  \end{tabular}
\right.
\end{equation}

This strongly enhanced energy dissipation via a moving ostacle close
to the transition might offer a practicable means of experimentally
testing our predictions \cite{PhysRevLett.85.2228}.

More generally at the qualitative level the strong coupling between the density and the spin mode approaching 
the ferromagnetic phase transition point should be reflected in a sudden emission of spin waves by exciting a density
modulation in the gas.

\section{Conclusion}

In conclusion, we have shown that two-component Bose gases present an
interesting scenario for the breaking of Goldstone modes. If the system
has a $U(1)\times\mathbf{Z}_2$ symmetry, the Goldstone mode related to the breaking of the global phase symmetry $U(1)$ in the 
condensed phase  
becomes not well defined at the critical point for the breaking the discrete symmetry $\mathbf{Z}_2$.  
When the system has  instead a  $U(1)\times U(1)\times \mathbf{Z}_2$ symmetry, the Goldstone mode related to the global 
phase (density mode) is strongly affected at the $\mathbf{Z}_2$ transition point, but still well defined in the limit of 
large wave lenghts. 
Although sometimes put on the same footing our results show even more that 2-component Bose-Einstein condensate 
with and without interconversion term are very different 

Let us here mention that our analysis can be extended to two and one dimensional systems, at least at the level of an 
effective low energy theory for mode coupling. The results are sketched in Table I. 
For a two dimensional gas a Belyaev analysis can be carried out without any problem.
For the density channel far for any instabilities the leading contribution is the same as for a single component Bose gas and 
it is proportional to $k^3$ (see, e.g., Ref. \cite{KopietzDecay}). For a mixture, i.e., $\Omega=0$, 
the decay rate at the phase separation point is bigger being proportional to  $k^2$, but still the phonons are well defined. 
Instead for $\Omega\neq 0$ at the ferromagnetic transition point one has a constant contribution at low momenta within 
Fermi's golden rule.

For a one dimensional gas some remarks are due.
First of all, the single component Bose gas is properly described by a Lieb-Liniger model. The system is integrable and 
therefore the modes do not decay.   
Our system is instead not integrable and therefore the density modes even far from any instability should have a 
finite life-time due to three density phonon processes. 
However the simple one-loop approximation failed in this case since energy and momentum conservation coincide. 
It was indeed first recognised by Andreev\cite{Andreev1D} and extended in the context of Luttinger liquid theory by 
Samokhin\cite{SamoLL}, that a more accurate analysis is required which leads to a decay rate proportional to $k^2$ 
(for a recent discussion see Ref.~\onlinecite{KopietzDecay}).     
On the other hand, for the decay of a density mode in two spin modes, the energy and the momentum conservation are distinct 
and therefore we can rely again on the one-loop analysis. We find that for a mixtures at the phase separation point the 
density mode decays as $k^{3/2}$, while it decays as $1/k$ at the ferromagnetic transition point when the interconversion term 
is present.
Although, as it is clear from the above discussion, in two and one dimension the perturbative analysis is not valid, 
it indicates, as expected, an increasingly strong effect in reduced dimensions on the density mode due to strong fluctuations 
of the spin density mode at criticality. 

Importantly, the effects here presented can be experimentally studied within present technology using trapped ultra-cold 
Bose gases with two hyperfine levels. The system has been indeed realised for the first time experimentally many years ago 
in the context of atom optics \cite{RabiCornell99,RabiCornell2000}, while the ferromagnetic-like transition has been more 
recently addressed in \cite{Zibold2010,KZOberthaler2015}. 
The main qualititauve signature being the emission of spin waves by perturbing the system via a density probe.

\acknowledgements
We thank Markus Oberthaler and Wilhelm Zwerger for useful
discussions. AR acknowledges support from the Alexander von Humboldt foundation and the hospitality of the condensed matter 
group at the TUM. FP acknowledges support by the APART 
fellowship of the Austrian Academy of Sciences.

\bibliography{phonon_decay}{}

\begin{thebibliography}{24}%
\makeatletter
\providecommand \@ifxundefined [1]{%
 \@ifx{#1\undefined}
}%
\providecommand \@ifnum [1]{%
 \ifnum #1\expandafter \@firstoftwo
 \else \expandafter \@secondoftwo
 \fi
}%
\providecommand \@ifx [1]{%
 \ifx #1\expandafter \@firstoftwo
 \else \expandafter \@secondoftwo
 \fi
}%
\providecommand \natexlab [1]{#1}%
\providecommand \enquote  [1]{``#1''}%
\providecommand \bibnamefont  [1]{#1}%
\providecommand \bibfnamefont [1]{#1}%
\providecommand \citenamefont [1]{#1}%
\providecommand \href@noop [0]{\@secondoftwo}%
\providecommand \href [0]{\begingroup \@sanitize@url \@href}%
\providecommand \@href[1]{\@@startlink{#1}\@@href}%
\providecommand \@@href[1]{\endgroup#1\@@endlink}%
\providecommand \@sanitize@url [0]{\catcode `\\12\catcode `\$12\catcode
  `\&12\catcode `\#12\catcode `\^12\catcode `\_12\catcode `\%12\relax}%
\providecommand \@@startlink[1]{}%
\providecommand \@@endlink[0]{}%
\providecommand \url  [0]{\begingroup\@sanitize@url \@url }%
\providecommand \@url [1]{\endgroup\@href {#1}{\urlprefix }}%
\providecommand \urlprefix  [0]{URL }%
\providecommand \Eprint [0]{\href }%
\providecommand \doibase [0]{http://dx.doi.org/}%
\providecommand \selectlanguage [0]{\@gobble}%
\providecommand \bibinfo  [0]{\@secondoftwo}%
\providecommand \bibfield  [0]{\@secondoftwo}%
\providecommand \translation [1]{[#1]}%
\providecommand \BibitemOpen [0]{}%
\providecommand \bibitemStop [0]{}%
\providecommand \bibitemNoStop [0]{.\EOS\space}%
\providecommand \EOS [0]{\spacefactor3000\relax}%
\providecommand \BibitemShut  [1]{\csname bibitem#1\endcsname}%
\let\auto@bib@innerbib\@empty
\bibitem [{\citenamefont {Goldstone}(1961)}]{Goldstone1961}%
  \BibitemOpen
  \bibfield  {author} {\bibinfo {author} {\bibfnamefont {J.}~\bibnamefont
  {Goldstone}},\ }\href {\doibase 10.1007/BF02812722} {\bibfield  {journal}
  {\bibinfo  {journal} {Il Nuovo Cimento (1955-1965)}\ }\textbf {\bibinfo
  {volume} {19}},\ \bibinfo {pages} {154} (\bibinfo {year} {1961})}\BibitemShut
  {NoStop}%
\bibitem [{\citenamefont {Englert}\ and\ \citenamefont
  {Brout}(1964)}]{Englert1964}%
  \BibitemOpen
  \bibfield  {author} {\bibinfo {author} {\bibfnamefont {F.}~\bibnamefont
  {Englert}}\ and\ \bibinfo {author} {\bibfnamefont {R.}~\bibnamefont
  {Brout}},\ }\href {\doibase 10.1103/PhysRevLett.13.321} {\bibfield  {journal}
  {\bibinfo  {journal} {Phys. Rev. Lett.}\ }\textbf {\bibinfo {volume} {13}},\
  \bibinfo {pages} {321} (\bibinfo {year} {1964})}\BibitemShut {NoStop}%
\bibitem [{\citenamefont {Higgs}(1964)}]{Higgs1964}%
  \BibitemOpen
  \bibfield  {author} {\bibinfo {author} {\bibfnamefont {P.~W.}\ \bibnamefont
  {Higgs}},\ }\href {\doibase 10.1103/PhysRevLett.13.508} {\bibfield  {journal}
  {\bibinfo  {journal} {Phys. Rev. Lett.}\ }\textbf {\bibinfo {volume} {13}},\
  \bibinfo {pages} {508} (\bibinfo {year} {1964})}\BibitemShut {NoStop}%
\bibitem [{\citenamefont {Anderson}(1958)}]{Anderson1958}%
  \BibitemOpen
  \bibfield  {author} {\bibinfo {author} {\bibfnamefont {P.~W.}\ \bibnamefont
  {Anderson}},\ }\href {\doibase 10.1103/PhysRev.110.827} {\bibfield  {journal}
  {\bibinfo  {journal} {Phys. Rev.}\ }\textbf {\bibinfo {volume} {110}},\
  \bibinfo {pages} {827} (\bibinfo {year} {1958})}\BibitemShut {NoStop}%
\bibitem [{\citenamefont {Wigner}(1938)}]{Wigner1938}%
  \BibitemOpen
  \bibfield  {author} {\bibinfo {author} {\bibfnamefont {E.}~\bibnamefont
  {Wigner}},\ }\href {\doibase 10.1039/TF9383400678} {\bibfield  {journal}
  {\bibinfo  {journal} {Trans. Faraday Soc.}\ }\textbf {\bibinfo {volume}
  {34}},\ \bibinfo {pages} {678} (\bibinfo {year} {1938})}\BibitemShut
  {NoStop}%
\bibitem [{\citenamefont {Pixley}\ \emph {et~al.}(2015)\citenamefont {Pixley},
  \citenamefont {Li},\ and\ \citenamefont {Das~Sarma}}]{DasSarma2015}%
  \BibitemOpen
  \bibfield  {author} {\bibinfo {author} {\bibfnamefont {J.~H.}\ \bibnamefont
  {Pixley}}, \bibinfo {author} {\bibfnamefont {X.}~\bibnamefont {Li}}, \ and\
  \bibinfo {author} {\bibfnamefont {S.}~\bibnamefont {Das~Sarma}},\ }\href
  {\doibase 10.1103/PhysRevLett.114.225303} {\bibfield  {journal} {\bibinfo
  {journal} {Phys. Rev. Lett.}\ }\textbf {\bibinfo {volume} {114}},\ \bibinfo
  {pages} {225303} (\bibinfo {year} {2015})}\BibitemShut {NoStop}%
\bibitem [{\citenamefont {Goldstein}\ and\ \citenamefont
  {Meystre}(1997)}]{Goldstein1997}%
  \BibitemOpen
  \bibfield  {author} {\bibinfo {author} {\bibfnamefont {E.~V.}\ \bibnamefont
  {Goldstein}}\ and\ \bibinfo {author} {\bibfnamefont {P.}~\bibnamefont
  {Meystre}},\ }\href {\doibase http://dx.doi.org/10.1103/PhysRevA.55.2935}
  {\bibfield  {journal} {\bibinfo  {journal} {Phys. Rev. A}\ }\textbf {\bibinfo
  {volume} {55}},\ \bibinfo {pages} {2935} (\bibinfo {year}
  {1997})}\BibitemShut {NoStop}%
\bibitem [{\citenamefont {Blakie}\ \emph {et~al.}(1999)\citenamefont {Blakie},
  \citenamefont {Ballagh},\ and\ \citenamefont {Gardiner}}]{Blakie1999}%
  \BibitemOpen
  \bibfield  {author} {\bibinfo {author} {\bibfnamefont {P.~B.}\ \bibnamefont
  {Blakie}}, \bibinfo {author} {\bibfnamefont {R.~J.}\ \bibnamefont {Ballagh}},
  \ and\ \bibinfo {author} {\bibfnamefont {C.~W.}\ \bibnamefont {Gardiner}},\
  }\href {\doibase 10.1088/1464-4266/1/4/304} {\bibfield  {journal} {\bibinfo
  {journal} {Journal of Optics B: Quantum and Semiclassical Optics}\ }\textbf
  {\bibinfo {volume} {1}},\ \bibinfo {pages} {378} (\bibinfo {year}
  {1999})}\BibitemShut {NoStop}%
\bibitem [{\citenamefont {Search}\ \emph {et~al.}()\citenamefont {Search},
  \citenamefont {Rojo},\ and\ \citenamefont {Berman}}]{Search2001}%
  \BibitemOpen
  \bibfield  {author} {\bibinfo {author} {\bibfnamefont {C.}~\bibnamefont
  {Search}}, \bibinfo {author} {\bibfnamefont {A.}~\bibnamefont {Rojo}}, \ and\
  \bibinfo {author} {\bibfnamefont {P.}~\bibnamefont {Berman}},\ }\href
  {\doibase 10.1103/PhysRevA.64.013615} {\bibinfo  {journal} {Phys. Rev. A}\ ,\
  \bibinfo {pages} {013615}}\BibitemShut {NoStop}%
\bibitem [{\citenamefont {Tommasini}\ \emph {et~al.}(2003)\citenamefont
  {Tommasini}, \citenamefont {Passos}, \citenamefont {Piza},\ and\
  \citenamefont {Hussein}}]{Tommasini2003}%
  \BibitemOpen
\bibfield  {journal} {  }\bibfield  {author} {\bibinfo {author} {\bibfnamefont
  {P.}~\bibnamefont {Tommasini}}, \bibinfo {author} {\bibfnamefont {E.~J.
  V.~D.}\ \bibnamefont {Passos}}, \bibinfo {author} {\bibfnamefont {A.~F. R.
  D.~T.}\ \bibnamefont {Piza}}, \ and\ \bibinfo {author} {\bibfnamefont
  {M.~S.}\ \bibnamefont {Hussein}},\ }\href {\doibase
  10.1103/PhysRevA.67.023606} {\bibfield  {journal} {\bibinfo  {journal} {Phys.
  Rev. A}\ }\textbf {\bibinfo {volume} {67}},\ \bibinfo {pages} {023606}
  (\bibinfo {year} {2003})}\BibitemShut {NoStop}%
\bibitem [{\citenamefont {Abad}\ and\ \citenamefont
  {Recati}(2013)}]{marta2013}%
  \BibitemOpen
  \bibfield  {author} {\bibinfo {author} {\bibfnamefont {M.}~\bibnamefont
  {Abad}}\ and\ \bibinfo {author} {\bibfnamefont {A.}~\bibnamefont {Recati}},\
  }\href {\doibase 10.1140/epjd/e2013-40053-2} {\bibfield  {journal} {\bibinfo
  {journal} {Eur. Phys. J D}\ }\textbf {\bibinfo {volume} {67}},\ \bibinfo
  {pages} {148} (\bibinfo {year} {2013})}\BibitemShut {NoStop}%
\bibitem [{\citenamefont {Zibold}\ \emph {et~al.}(2010)\citenamefont {Zibold},
  \citenamefont {Nicklas}, \citenamefont {Gross},\ and\ \citenamefont
  {Oberthaler}}]{Zibold2010}%
  \BibitemOpen
  \bibfield  {author} {\bibinfo {author} {\bibfnamefont {T.}~\bibnamefont
  {Zibold}}, \bibinfo {author} {\bibfnamefont {E.}~\bibnamefont {Nicklas}},
  \bibinfo {author} {\bibfnamefont {C.}~\bibnamefont {Gross}}, \ and\ \bibinfo
  {author} {\bibfnamefont {M.~K.}\ \bibnamefont {Oberthaler}},\ }\href
  {\doibase 10.1103/PhysRevLett.105.204101} {\bibfield  {journal} {\bibinfo
  {journal} {Phys. Rev. Lett.}\ }\textbf {\bibinfo {volume} {105}},\ \bibinfo
  {pages} {204101} (\bibinfo {year} {2010})}\BibitemShut {NoStop}%
\bibitem [{\citenamefont {Landau}\ and\ \citenamefont {Lifshitz}(1980)}]{LL9}%
  \BibitemOpen
  \bibfield  {author} {\bibinfo {author} {\bibfnamefont {L.~D.}\ \bibnamefont
  {Landau}}\ and\ \bibinfo {author} {\bibfnamefont {E.~M.}\ \bibnamefont
  {Lifshitz}},\ }\href@noop {} {\emph {\bibinfo {title} {Statistical Physics
  part. 2}}}\ (\bibinfo  {publisher} {Pergamon Press},\ \bibinfo {year}
  {1980})\BibitemShut {NoStop}%
\bibitem [{\citenamefont {Stenger}\ \emph {et~al.}(1999)\citenamefont
  {Stenger}, \citenamefont {Inouye}, \citenamefont {Chikkatur}, \citenamefont
  {Stamper-Kurn}, \citenamefont {Pritchard},\ and\ \citenamefont
  {Ketterle}}]{KetterleBragg}%
  \BibitemOpen
  \bibfield  {author} {\bibinfo {author} {\bibfnamefont {J.}~\bibnamefont
  {Stenger}}, \bibinfo {author} {\bibfnamefont {S.}~\bibnamefont {Inouye}},
  \bibinfo {author} {\bibfnamefont {A.~P.}\ \bibnamefont {Chikkatur}}, \bibinfo
  {author} {\bibfnamefont {D.~M.}\ \bibnamefont {Stamper-Kurn}}, \bibinfo
  {author} {\bibfnamefont {D.~E.}\ \bibnamefont {Pritchard}}, \ and\ \bibinfo
  {author} {\bibfnamefont {W.}~\bibnamefont {Ketterle}},\ }\href {\doibase
  10.1103/PhysRevLett.82.4569} {\bibfield  {journal} {\bibinfo  {journal}
  {Phys. Rev. Lett.}\ }\textbf {\bibinfo {volume} {82}},\ \bibinfo {pages}
  {4569} (\bibinfo {year} {1999})}\BibitemShut {NoStop}%
\bibitem [{\citenamefont {Steinhauer}\ \emph {et~al.}(2002)\citenamefont
  {Steinhauer}, \citenamefont {Ozeri}, \citenamefont {Katz},\ and\
  \citenamefont {Davidson}}]{SteinhauerSpec}%
  \BibitemOpen
  \bibfield  {author} {\bibinfo {author} {\bibfnamefont {J.}~\bibnamefont
  {Steinhauer}}, \bibinfo {author} {\bibfnamefont {R.}~\bibnamefont {Ozeri}},
  \bibinfo {author} {\bibfnamefont {N.}~\bibnamefont {Katz}}, \ and\ \bibinfo
  {author} {\bibfnamefont {N.}~\bibnamefont {Davidson}},\ }\href {\doibase
  10.1103/PhysRevLett.88.120407} {\bibfield  {journal} {\bibinfo  {journal}
  {Phys. Rev. Lett.}\ }\textbf {\bibinfo {volume} {88}},\ \bibinfo {pages}
  {120407} (\bibinfo {year} {2002})}\BibitemShut {NoStop}%
\bibitem [{\citenamefont {{Landig Renate}}\ \emph {et~al.}(2015)\citenamefont
  {{Landig Renate}}, \citenamefont {{Brennecke Ferdinand}}, \citenamefont
  {{Mottl Rafael}}, \citenamefont {{Donner Tobias}},\ and\ \citenamefont
  {{Esslinger Tilman}}}]{EsslingerSkw}%
  \BibitemOpen
  \bibfield  {author} {\bibinfo {author} {\bibnamefont {{Landig Renate}}},
  \bibinfo {author} {\bibnamefont {{Brennecke Ferdinand}}}, \bibinfo {author}
  {\bibnamefont {{Mottl Rafael}}}, \bibinfo {author} {\bibnamefont {{Donner
  Tobias}}}, \ and\ \bibinfo {author} {\bibnamefont {{Esslinger Tilman}}},\
  }\href {\doibase http://dx.doi.org/10.1038/ncomms8046 10.1038/ncomms8046}
  {\bibfield  {journal} {\bibinfo  {journal} {Nat Commun}\ }\textbf {\bibinfo
  {volume} {6}} (\bibinfo {year} {2015}),\ http://dx.doi.org/10.1038/ncomms8046
  10.1038/ncomms8046},\ \bibinfo {note} {supplementary information available
  for this article at
  http://www.nature.com/ncomms/2015/150506/ncomms8046/suppinfo/ncomms8046\_S1.html}\BibitemShut
  {NoStop}%
\bibitem [{\citenamefont {Carusotto}\ \emph {et~al.}(2006)\citenamefont
  {Carusotto}, \citenamefont {Hu}, \citenamefont {Collins},\ and\ \citenamefont
  {Smerzi}}]{PhysRevLett.97.260403}%
  \BibitemOpen
  \bibfield  {author} {\bibinfo {author} {\bibfnamefont {I.}~\bibnamefont
  {Carusotto}}, \bibinfo {author} {\bibfnamefont {S.~X.}\ \bibnamefont {Hu}},
  \bibinfo {author} {\bibfnamefont {L.~A.}\ \bibnamefont {Collins}}, \ and\
  \bibinfo {author} {\bibfnamefont {A.}~\bibnamefont {Smerzi}},\ }\href
  {\doibase 10.1103/PhysRevLett.97.260403} {\bibfield  {journal} {\bibinfo
  {journal} {Phys. Rev. Lett.}\ }\textbf {\bibinfo {volume} {97}},\ \bibinfo
  {pages} {260403} (\bibinfo {year} {2006})}\BibitemShut {NoStop}%
\bibitem [{\citenamefont {Onofrio}\ \emph {et~al.}(2000)\citenamefont
  {Onofrio}, \citenamefont {Raman}, \citenamefont {Vogels}, \citenamefont
  {Abo-Shaeer}, \citenamefont {Chikkatur},\ and\ \citenamefont
  {Ketterle}}]{PhysRevLett.85.2228}%
  \BibitemOpen
  \bibfield  {author} {\bibinfo {author} {\bibfnamefont {R.}~\bibnamefont
  {Onofrio}}, \bibinfo {author} {\bibfnamefont {C.}~\bibnamefont {Raman}},
  \bibinfo {author} {\bibfnamefont {J.~M.}\ \bibnamefont {Vogels}}, \bibinfo
  {author} {\bibfnamefont {J.~R.}\ \bibnamefont {Abo-Shaeer}}, \bibinfo
  {author} {\bibfnamefont {A.~P.}\ \bibnamefont {Chikkatur}}, \ and\ \bibinfo
  {author} {\bibfnamefont {W.}~\bibnamefont {Ketterle}},\ }\href {\doibase
  10.1103/PhysRevLett.85.2228} {\bibfield  {journal} {\bibinfo  {journal}
  {Phys. Rev. Lett.}\ }\textbf {\bibinfo {volume} {85}},\ \bibinfo {pages}
  {2228} (\bibinfo {year} {2000})}\BibitemShut {NoStop}%
\bibitem [{\citenamefont {Lange}\ \emph {et~al.}(2012)\citenamefont {Lange},
  \citenamefont {Kopietz},\ and\ \citenamefont {Kreisel}}]{KopietzDecay}%
  \BibitemOpen
  \bibfield  {author} {\bibinfo {author} {\bibfnamefont {P.}~\bibnamefont
  {Lange}}, \bibinfo {author} {\bibfnamefont {P.}~\bibnamefont {Kopietz}}, \
  and\ \bibinfo {author} {\bibfnamefont {A.}~\bibnamefont {Kreisel}},\ }\href
  {\doibase 10.1140/epjb/e2012-30639-3} {\bibfield  {journal} {\bibinfo
  {journal} {The European Physical Journal B}\ }\textbf {\bibinfo {volume}
  {85}},\ \bibinfo {pages} {1} (\bibinfo {year} {2012})}\BibitemShut {NoStop}%
\bibitem [{\citenamefont {Andreev}(1980)}]{Andreev1D}%
  \BibitemOpen
  \bibfield  {author} {\bibinfo {author} {\bibfnamefont {A.~F.}\ \bibnamefont
  {Andreev}},\ }\href@noop {} {\bibfield  {journal} {\bibinfo  {journal} {Sov.
  Phys. -- JETP}\ }\textbf {\bibinfo {volume} {51}},\ \bibinfo {pages} {1038}
  (\bibinfo {year} {1980})}\BibitemShut {NoStop}%
\bibitem [{\citenamefont {Samokhin}(1998)}]{SamoLL}%
  \BibitemOpen
  \bibfield  {author} {\bibinfo {author} {\bibfnamefont {K.~V.}\ \bibnamefont
  {Samokhin}},\ }\href@noop {} {\bibfield  {journal} {\bibinfo  {journal} {J.
  Phys.: Condens. Mater.}\ }\textbf {\bibinfo {volume} {10}},\ \bibinfo {pages}
  {L533} (\bibinfo {year} {1998})}\BibitemShut {NoStop}%
\bibitem [{\citenamefont {Matthews}\ \emph {et~al.}(1999)\citenamefont
  {Matthews}, \citenamefont {Anderson}, \citenamefont {Haljan}, \citenamefont
  {Hall}, \citenamefont {Holland}, \citenamefont {Williams}, \citenamefont
  {Wieman},\ and\ \citenamefont {Cornell}}]{RabiCornell99}%
  \BibitemOpen
  \bibfield  {author} {\bibinfo {author} {\bibfnamefont {M.~R.}\ \bibnamefont
  {Matthews}}, \bibinfo {author} {\bibfnamefont {B.~P.}\ \bibnamefont
  {Anderson}}, \bibinfo {author} {\bibfnamefont {P.~C.}\ \bibnamefont
  {Haljan}}, \bibinfo {author} {\bibfnamefont {D.~S.}\ \bibnamefont {Hall}},
  \bibinfo {author} {\bibfnamefont {M.~J.}\ \bibnamefont {Holland}}, \bibinfo
  {author} {\bibfnamefont {J.~E.}\ \bibnamefont {Williams}}, \bibinfo {author}
  {\bibfnamefont {C.~E.}\ \bibnamefont {Wieman}}, \ and\ \bibinfo {author}
  {\bibfnamefont {E.~A.}\ \bibnamefont {Cornell}},\ }\href {\doibase
  10.1103/PhysRevLett.83.3358} {\bibfield  {journal} {\bibinfo  {journal}
  {Phys. Rev. Lett.}\ }\textbf {\bibinfo {volume} {83}},\ \bibinfo {pages}
  {3358} (\bibinfo {year} {1999})}\BibitemShut {NoStop}%
\bibitem [{\citenamefont {Williams}\ \emph {et~al.}(2000)\citenamefont
  {Williams}, \citenamefont {Walser}, \citenamefont {Cooper}, \citenamefont
  {Cornell},\ and\ \citenamefont {Holland}}]{RabiCornell2000}%
  \BibitemOpen
  \bibfield  {author} {\bibinfo {author} {\bibfnamefont {J.}~\bibnamefont
  {Williams}}, \bibinfo {author} {\bibfnamefont {R.}~\bibnamefont {Walser}},
  \bibinfo {author} {\bibfnamefont {J.}~\bibnamefont {Cooper}}, \bibinfo
  {author} {\bibfnamefont {E.~A.}\ \bibnamefont {Cornell}}, \ and\ \bibinfo
  {author} {\bibfnamefont {M.}~\bibnamefont {Holland}},\ }\href {\doibase
  10.1103/PhysRevA.61.033612} {\bibfield  {journal} {\bibinfo  {journal} {Phys.
  Rev. A}\ }\textbf {\bibinfo {volume} {61}},\ \bibinfo {pages} {033612}
  (\bibinfo {year} {2000})}\BibitemShut {NoStop}%
\bibitem [{\citenamefont {Nicklas}\ \emph {et~al.}(2015)\citenamefont
  {Nicklas}, \citenamefont {Karl}, \citenamefont {H\"ofer}, \citenamefont
  {Johnson}, \citenamefont {Muessel}, \citenamefont {Strobel}, \citenamefont
  {Tomkovi\ifmmode~\check{c}\else \v{c}\fi{}}, \citenamefont {Gasenzer},\ and\
  \citenamefont {Oberthaler}}]{KZOberthaler2015}%
  \BibitemOpen
  \bibfield  {author} {\bibinfo {author} {\bibfnamefont {E.}~\bibnamefont
  {Nicklas}}, \bibinfo {author} {\bibfnamefont {M.}~\bibnamefont {Karl}},
  \bibinfo {author} {\bibfnamefont {M.}~\bibnamefont {H\"ofer}}, \bibinfo
  {author} {\bibfnamefont {A.}~\bibnamefont {Johnson}}, \bibinfo {author}
  {\bibfnamefont {W.}~\bibnamefont {Muessel}}, \bibinfo {author} {\bibfnamefont
  {H.}~\bibnamefont {Strobel}}, \bibinfo {author} {\bibfnamefont
  {J.}~\bibnamefont {Tomkovi\ifmmode~\check{c}\else \v{c}\fi{}}}, \bibinfo
  {author} {\bibfnamefont {T.}~\bibnamefont {Gasenzer}}, \ and\ \bibinfo
  {author} {\bibfnamefont {M.~K.}\ \bibnamefont {Oberthaler}},\ }\href
  {\doibase 10.1103/PhysRevLett.115.245301} {\bibfield  {journal} {\bibinfo
  {journal} {Phys. Rev. Lett.}\ }\textbf {\bibinfo {volume} {115}},\ \bibinfo
  {pages} {245301} (\bibinfo {year} {2015})}\BibitemShut {NoStop}%
\end{thebibliography}%

\end{document}